# Noninvasive In-vivo Estimation of HbA1c Based on Beer–Lambert Model from Photoplethysmogram Using Only Two Wavelengths

Mrinmoy Sarker Turja [1], Tae-Ho Kwon[1], Hyoungkeun Kim[2] and Ki-Doo Kim[1,*]

[1] Department of Electronics Engineering, Kookmin University, Seoul 02707, Korea
[2] Korea I.T.S. Co., Ltd., Gangnam, Seoul 06373, Korea
*Correspondence: kdk@kookmin.ac.kr (K.-D.K)

**Abstract:** Glycated hemoglobin (HbA1c) is the most important factor in diabetes control. Since HbA1c reflects the average blood glucose level over the preceding three months, it is unaffected by the patient's activity level or diet before the test. Noninvasive HbA1c measurement reduces both the pain and complications associated with fingertip piercing to collect blood. Photoplethysmography is helpful for measuring HbA1c without blood samples. Herein, only two wavelengths (615 and 525 nm) were used to estimate HbA1c noninvasively, where two different ratio calibrations were applied and performances were compared to a work that uses three wavelengths. For the fingertip type, the Pearson's r values for HbA1c estimates are 0.896 and 0.905 considering ratio calibrations for blood-vessel and whole-finger models, respectively. Using another value (HbA1c) calibration in addition to ratio calibrations, we can improve this performance, such that the Pearson's r values of HbA1c levels are 0.929 and 0.930 for blood-vessel and whole-finger models, respectively. In the previous study using three wavelengths, the Pearson's r values were 0.916 and 0.959 for the blood-vessel and whole-finger models, respectively. Here, the RCF of $SpO_2$ estimation is 0.986 when $SpO_2$ ratio calibration is applied, while in the previous study, the RCF values of $SpO_2$ estimation were 0.983 and 0.986 for the blood-vessel and whole-finger models, respectively. Thus, we show that HbA1c estimation using only two wavelengths has comparable performance to previous studies.

## 1. Introduction

Traditional blood glucose testing often requires blood samples, which can be uncomfortable for the patient and increase the risk of skin and red-blood-cell-life abnormalities. On the other hand, the glycated hemoglobin (HbA1c) value reflects the average blood glucose level over the previous three months, and it is unaffected by physical activity or food intake for several hours leading up to the measurement. High levels of HbA1c indicate poor blood glucose control. As a result, it is used as the most fundamental indicator of the degree of blood glucose control over a period and to predict the onset of long-term issues due to diabetes. Human blood consists of 55% yellow liquid plasma, 44% solid red blood cells, and 1% white blood cells and platelets. Hemoglobin is a type of protein found in the red blood cells that plays an important role in transporting oxygen by binding to the oxygen in the blood and glucose contained in blood cells. Hemoglobin bound to glucose is called glycated hemoglobin. Regarding glycated hemoglobin, the more glucose there is in the blood, the more the hemoglobin in the red blood cells binds to the glucose, resulting in higher blood glucose levels. The glycated hemoglobin level is the ratio of glycated hemoglobin to total hemoglobin in the blood. Because the normal lifespan of red blood cells is about 4 months and the lifespan of individual red blood cells can vary widely, this test can only provide the HbA1c estimate over the preceding 3 months. People with diabetes are at increased risk of developing additional medical complications, such as heart disease, kidney failure, stroke, cataracts, and/or premature death. Therefore, early diagnosis of diabetes in the prediabetic stage is crucial for preventing the deterioration of the blood-sugar regulation system. This can be achieved through various blood-based tests, such as random, oral, and fasting blood glucose tests or glycated hemoglobin (HbA1c) tests. These tests are used to detect the levels of glucose in the blood, which is a key indicator of diabetes. In diagnosing diabetes, the HbA1c test is known to have better performance than the plasma glucose test [1].

Many enzymatic and nonenzymatic electrochemical glucose sensors [2–7] have been created over the past few decades; however, these approaches are invasive. Immunoassay, ion-exchange high-performance liquid chromatography (HPLC), boronated affinity chromatography, and capillary electrophoresis (CE) are the four most commonly used methods for estimating HbA1c [8]. Both HPLC-electrospray mass spectrometry (HPLC-ESI/MS) and HPLC-capillary electrophoresis-ultraviolet (HPLC-CE-UV) are recommended by the International Federation on Clinical Chemistry and Laboratory Medicine (IFCC) for measuring HbA1c level in human blood [9]. All of these methods still require blood samples. However, the development and practical use of noninvasive HbA1c estimation have been of increasing interest recently. One study [10] only addressed photoplethysmography (PPG) sensor design and did not consider noninvasive in-vivo estimation techniques while discussing estimation of in-vitro HbA1c levels. Based on the measurement conditions related to hyperglycemia, researchers have divided mouse models into diabetic, obese, and normal control categories [11]. In [12], the authors reported that individuals could be divided into diabetic and non-diabetic categories using PPG signals. In another study [13], the HbA1c levels were calculated by also taking into account the acetone levels in the breath. In our previous study [14], we estimated HbA1c noninvasively using Beer–Lambert-law-based model with three wavelengths. However, the three wavelengths used made the model a bit complicated and inconvenient for the user as well. To address these issues, a noninvasive estimation strategy using PPG signals with only two wavelengths is proposed herein. In this study, we used a white LED to provide signals at three wavelengths (465, 525, and 615 nm); of these, red (615 nm) and green (525 nm) are selected to prove the proposed method. Although any two of the three wavelengths can be used, the essence of this study does not change, and the detailed explanation for choosing the wavelength pair (525, 615 nm) is given in Appendix A.

## 2. Background

*2.1. Beer–Lambert Law*

Beer–Lambert law specifies the attenuation of light passing through a sample [15]. In most cases, the Beer–Lambert law is suitable for quantifying the concentration of a compound remaining in a sample. Accordingly, the attenuation of light is directly proportional to the concentration of the residual compounds in the sample. The practical expression of the Beer–Lambert law is given in Equation (1).

$$A = \varepsilon \times c \times d \quad (1)$$

where A is the total absorption, $\varepsilon$ is the molar absorption coefficient (L·mol$^{-1}$·cm$^{-1}$), $c$ is the concentration of the attenuating species (mol·cm$^{-1}$), and $d$ is the optical path length (cm). Equation (1) can also be expressed in terms of the incident light intensity on the sample and transmitted light intensity through the sample.

$$A = \log \frac{I_0}{I} \qquad (2)$$

where $I_0$ is the intensity of light incident on the sample, and $I$ is the intensity of the transmitted light through the sample.

*2.2. Finger-Type Models: Blood-Vessel and Whole-Finger Models*

2.2.1. Blood-vessel model

The blood-vessel model was created using the assumption that the diameter of the vessel increases slightly to accommodate the volume of blood as it enters the vessel and decreases as the blood leaves the channel. This is illustrated in Figure 1.

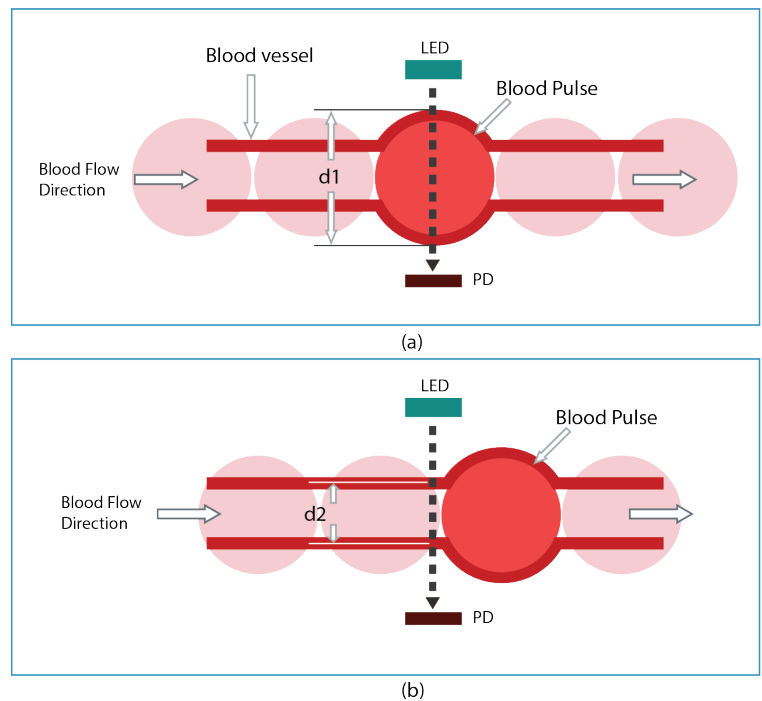

**Figure 1.** Blood-vessel model: (a) light intensity in the systolic phase, and (**b**) light intensity in the diastolic phase.

Considering Equation (1), blood can be defined as a homogeneous solution of *HbA1c*, *HbO*, and *HHb*. Hence, the total absorption at wavelength $\lambda$ can be expressed as

$$A = (\varepsilon_{HbA1c}(\lambda) \times c_{HbA1c} + \varepsilon_{HbO}(\lambda) \times c_{HbO} + \varepsilon_{HHb}(\lambda) \times c_{HHb}) \times d \qquad (3)$$

where $\varepsilon_{HbO}(\lambda), \varepsilon_{HHb}(\lambda), and\ \varepsilon_{HbA1c}(\lambda)$ are the molar absorption coefficients at wavelength $\lambda$ for *HbO* (oxygenated hemoglobin), *HHb* (deoxygenated hemoglobin), and *HbA1c*, respectively; $c$ represents the molar concentration of each element, and $d$ is the distance traveled by light.

%SpO$_2$ and %HbA1c can be described as follows:

$$\%SpO_2 = \frac{c_{HbO}}{c_{HbO} + c_{HHb}} \times 100\% \qquad (4)$$

$$\%HbA1c = P_{HbA1c} \times 100\% \qquad (5)$$

The partial molar concentrations of *HbO*, *HHb*, and *HbA1c* can be expressed as *P*$_{HbO}$, *P*$_{HHb}$, and *P*$_{HbA1c}$, respectively, as follows:

$$P_{HbO} = \frac{c_{HbO}}{c_T} \tag{6}$$

$$P_{HHb} = \frac{c_{HHb}}{c_T} \tag{7}$$

$$P_{HbA1c} = \frac{c_{HbA1c}}{c_T} \tag{8}$$

$$c_T = c_{HbO} + c_{HHb} + c_{HbA1c} \tag{9}$$

From Equations (6) to (9),

$$P_{HHb} = 1 - (P_{HbO} + P_{HbA1c}) \tag{10}$$

Equation (4) can be expressed in terms of the partial molar concentration.

$$\%SpO_2 = \frac{P_{Hbo}}{P_{HbO} + P_{HHb}} \times 100\% \tag{11}$$

When the blood vessel expands, Equation (3) becomes

$$\Delta A = (\varepsilon_{HbO}(\lambda) \times c_{HbO} + \varepsilon_{HHb}(\lambda) \times c_{HHb} + \varepsilon_{HbA1c}(\lambda) \times c_{HbA1c}) \times \Delta d \tag{12}$$

where $\Delta A = A1 - A2$, $\Delta d = d_1 - d_2$; $A1$ represents the absorbance when blood enters the vessel, and $A2$ represents the absorbance when the blood flows out from the vessel. The variables $d_1$ and $d_2$ represent the diameters of the blood vessel as blood enters and leaves the vessel, respectively.

For the two wavelengths considered in this study (i.e., $\lambda 1 = 525$ nm and $\lambda 2 = 615$ nm), Equation (12) can be expressed as

$$\Delta A_{\lambda 1} = (\varepsilon_{HbO}(\lambda 1) \times c_{HbO} + \varepsilon_{HHb}(\lambda 1) \times c_{HHb} + \varepsilon_{HbA1c}(\lambda 1) \times c_{HbA1c}) \times \Delta d \tag{13}$$

$$\Delta A_{\lambda 2} = (\varepsilon_{HbO}(\lambda 2) \times c_{HbO} + \varepsilon_{HHb}(\lambda 2) \times c_{HHb} + \varepsilon_{HbA1c}(\lambda 2) \times c_{HbA1c}) \times \Delta d \tag{14}$$

From Equations (13) and (14), the ratio equation can be obtained to estimate the unknown parameter $P_{HbA1c}$. The ratio equation can be expressed as

$$R = \frac{\Delta A_{\lambda 2}}{\Delta A_{\lambda 1}} = \frac{(\varepsilon_{HbO}(\lambda 2) \times c_{HbO} + \varepsilon_{HHb}(\lambda 2) \times c_{HHb} + \varepsilon_{HbA1c}(\lambda 2) \times c_{HbA1c}) \times \Delta d}{(\varepsilon_{HbO}(\lambda 1) \times c_{HbO} + \varepsilon_{HHb}(\lambda 1) \times c_{HHb} + \varepsilon_{HbA1c}(\lambda 1) \times c_{HbA1c}) \times \Delta d} \tag{15}$$

Replacing the molar concentration with the partial molar concentration, we get

$$R = \frac{\Delta A_{\lambda 2}}{\Delta A_{\lambda 1}} = \frac{(\varepsilon_{HbO}(\lambda 2) \times P_{HbO} + \varepsilon_{HHb}(\lambda 2) \times P_{HHb} + \varepsilon_{HbA1c}(\lambda 2) \times P_{HbA1c})}{(\varepsilon_{HbO}(\lambda 1) \times P_{HbO} + \varepsilon_{HHb}(\lambda 1) \times P_{HHb} + \varepsilon_{HbA1c}(\lambda 1) \times P_{HbA1c})} \tag{16}$$

Equations (13) and (14) can be expressed in the form of Equation (2) as follows:

$$\Delta A_{\lambda 1} = \Delta \left[\log \frac{I_0}{I}\right]_{\lambda 1} \tag{17}$$

$$\Delta A_{\lambda 2} = \Delta \left[\log \frac{I_0}{I}\right]_{\lambda 2} \tag{18}$$

Now, Equation (16) can be expressed by combining Equations (17) and (18). Therefore, the ratio can be calculated directly from the light received at the fingertip.

$$R = \frac{\Delta\left[\log\frac{I_0}{I}\right]_{\lambda 2}}{\Delta\left[\log\frac{I_0}{I}\right]_{\lambda 1}} = \frac{\left[\log\frac{I_0(d1)}{I(d1)} - \log\frac{I_0(d2)}{I(d2)}\right]_{\lambda 2}}{\left[\log\frac{I_0(d1)}{I(d1)} - \log\frac{I_0(d2)}{I(d2)}\right]_{\lambda 1}} = \frac{\left[\log\frac{I(d1)}{I(d2)}\right]_{\lambda 2}}{\left[\log\frac{I(d1)}{I(d2)}\right]_{\lambda 1}} \quad (19)$$

Using Equations (10) and (11), we get

$$P_{HbO} = SpO_2 * (1 - P_{HbA1c}) \quad (20)$$

Equation (10) can be expressed in terms of $SpO_2$ and $P_{HbA1c}$ as

$$\begin{aligned}P_{HHb} &= 1 - (P_{HbO} + P_{HbA1c}) \\ &= (1 - SpO_2) * (1 - P_{HbA1c})\end{aligned} \quad (21)$$

Solving for the values of $P_{HbO}$ and $P_{HHb}$, the following equation is obtained:

$$R = \frac{P_{HbA1c}[\varepsilon_{HbA1c}(\lambda 2) - \varepsilon_{HHb}(\lambda 2) \times (1 - SpO_2) - \varepsilon_{HbO}(\lambda 2) \times SpO_2] + [\varepsilon_{HHb}(\lambda 2) \times (1 - SpO_2) + \varepsilon_{HbO}(\lambda 2) \times SpO_2]}{P_{HbA1c}[\varepsilon_{HbA1c}(\lambda 1) - \varepsilon_{HHb}(\lambda 1) \times (1 - SpO_2) - \varepsilon_{HbO}(\lambda 1) \times SpO_2] + [\varepsilon_{HHb}(\lambda 1) \times (1 - SpO_2) + \varepsilon_{HbO}(\lambda 1) \times SpO_2]} \quad (22)$$

or

$$P_{HbA1c} = \frac{[\varepsilon_{HHb}(\lambda 2) \times (1 - SpO_2) + \varepsilon_{HbO}(\lambda 2) \times SpO_2] - R[\varepsilon_{HHb}(\lambda 1) \times (1 - SpO_2) + \varepsilon_{HbO}(\lambda 1) \times SpO_2]}{R[\varepsilon_{HbA1c}(\lambda 1) - \varepsilon_{HHb}(\lambda 1) \times (1 - SpO_2) - \varepsilon_{HbO}(\lambda 1) \times SpO_2] - [\varepsilon_{HbA1c}(\lambda 2) - \varepsilon_{HHb}(\lambda 2) \times (1 - SpO_2) - \varepsilon_{HbO}(\lambda 2) \times SpO_2]} \quad (23)$$

The values of the molar absorption coefficients of *HbA1c*, *HbO*, and *HHb* for two different wavelengths (525 and 615 nm) are given in Table 1. The molar absorption coefficients of *HbA1c* were taken from Hossain et al. [16] and those of *HbO* and *HHb* were considered from [17].

**Table 1.** Molar absorption coefficients for the blood-vessel model.

| Substance | Molar absorption coefficient (cm$^{-1}$·M$^{-1}$) | |
|---|---|---|
| | $\lambda 1 = 525$ nm | $\lambda 2 = 615$ nm |
| *HbA1c* | 455,139.5677 | 170,5554218 |
| *HbO* | 30,882.8 | 1166.4 |
| *HHb* | 35,170.8 | 7553.4 |

2.2.2. Whole-finger model

The whole-finger model considers the lumped fingertip constituents as a homogeneous mixture. The blood entering this model will increase the fractional volume of arterial blood. Considering only the increase in the arterial fraction, the equation for the absorption coefficient becomes [14].

$$C_a = \left(V_a \mu_a^{art}(\lambda) + V_v \mu_a^{vein}(\lambda) + V_w \mu_a^{water}(\lambda) + (1 - (V_a + V_v + V_w)) \times \mu_a^{baseline}\right) \quad (24)$$

Equations (25) and (26) can be easily obtained [14] after replacing the values of $P_{HbO}$ and $P_{HHb}$ from Equations (20) and (21).

$$\mu_a^{art} = \{P_{HbA1c}(\mu_a^{HbA1c}(\lambda) - \mu_a^{HbO}(\lambda)SpO2 - \mu_a^{HHb}(\lambda)(1 - SpO2)) + \mu_a^{HbO}(\lambda)SpO2 + \mu_a^{HHb}(\lambda)(1 - SpO2)\} \quad (25)$$

$$\mu_a^{vein} = \{P_{HbA1c}(\mu_a^{HbA1c}(\lambda) - \mu_a^{HbO}(\lambda)SpO2 - \mu_a^{HHb}(\lambda)(1 - SpO2)) + \mu_a^{HbO}(\lambda)SpO2 + \mu_a^{HHb}(\lambda)(1 - SpO2)\} \quad (26)$$

Now, considering the arterial fraction increment, the absorption coefficient equation is as shown in Equation (27). Here, $\Delta C_a$ represents the change in the absorption coefficient due to a change in the arterial blood volume.

$$C_a + \Delta C_a = \left((V_a + \Delta V_a)\mu_a^{art}(\lambda) + V_v \mu_a^{vein}(\lambda) + V_w \mu_a^{water}(\lambda) + (1 - (V_a + \Delta V_a + V_v + V_w)) \times \mu_a^{baseline}\right) \quad (27)$$

After subtracting Equation (24) from (27), we get

$$\Delta C_a = \Delta V_a (\mu_a^{art}(\lambda) - \mu_a^{baseline}(\lambda)) \tag{28}$$

From Beer–Lambert law, we get

$$I = I_o \times 10^{-C_a d} \tag{29}$$

Equation (29) needs to be differentiated in terms of $C_a$ to find the relationship between the physical light intensity and Equation (28).

$$\frac{dI}{dC_a} = -\ln(10) I_o d \times 10^{-C_a d} \tag{30}$$

$$\frac{dI}{dC_a} \approx \frac{\Delta I}{\Delta C_a} \tag{31}$$

From Equations (30) and (31), we get

$$\Delta I \approx -\ln(10) I_o \Delta C_a d \times 10^{-C_a d} \tag{32}$$

Now, the AC–DC intensity ratio is generated by the assumption $\frac{I_{AC}}{I_{DC}} = \frac{\Delta I}{I}$. Hence, dividing Equation (32) by Equation (29) gives

$$\frac{\Delta I}{I} \approx -\ln(10) \Delta V_a \left(\mu_a^{art}(\lambda) - \mu_a^{baseline}(\lambda)\right) d \tag{33}$$

Thus, the ratio equation becomes

$$R = \frac{\left[\frac{\Delta I}{I}\right]_{\lambda_2}}{\left[\frac{\Delta I}{I}\right]_{\lambda_1}} = \frac{\mu_a^{art}(\lambda 2) - \mu_a^{baseline}(\lambda 2)}{\mu_a^{art}(\lambda 1) - \mu_a^{baseline}(\lambda 1)} \tag{34}$$

After solving Equation (34) for $P_{HbA1c}$, we get Equation (35).

$$P_{HbA1c} = \frac{\mu_a^{HbO}(\lambda 2) \cdot SpO2 + \mu_a^{HHb}(\lambda 2) \cdot (1-SpO2) - \mu_a^{baseline}(\lambda 2) - R \cdot \left(\mu_a^{HbO}(\lambda 2) \cdot SpO2 + \mu_a^{HHb}(\lambda 2) \cdot (1-SpO2) - \mu_a^{baseline}(\lambda 2)\right)}{R \cdot \left(\mu_a^{HbA1c}(\lambda 1) - \mu_a^{HbO}(\lambda 1) \cdot SpO2 - \mu_a^{HHb}(\lambda 1) \cdot (1-SpO2)\right) - \left(\mu_a^{HbA1c}(\lambda 1) - \mu_a^{HbO}(\lambda 1) \cdot SpO2 - \mu_a^{HHb}(\lambda 1) \cdot (1-SpO2)\right)} \tag{35}$$

The values of the molar absorption coefficients of *HbA1c*, *HbO*, *HHb*, skin baseline, and water for the two wavelengths (525 and 615 nm) are given in Table 2. The molar absorption coefficients of *HbA1c*, *HbO*, and *HHb* were considered from the study mentioned before, and the skin baseline and water values were considered from Saidi [18] and Segelstein [19], respectively.

**Table 2.** Absorption coefficients for the whole-finger model.

| Substance | Absorption coefficient (cm⁻¹) | |
|---|---|---|
| | $\lambda 1 = 525$ nm | $\lambda 2 = 615$ nm |
| *HbA1c* | 1058.4641 | 396.6405 |
| *HbO* | 71.8205 | 2.7126 |
| *HHb* | 81.7926 | 17.566 |
| Skin Baseline | 1.0966 | 0.6552 |
| Water | 0.0003927 | 0.0027167 |

### 2.3. $SpO_2$ Calculation

To calculate the $SpO_2$ values from the PPG signals, we followed the method in [20]. The ratio $R_{SpO_2}$ was calculated from the ratio of the normalized intensity of the received green light ($I_{n_{\lambda_1}}$) to red light ($I_{n_{\lambda_2}}$), which is expressed as Equation (24).

$$R_{SpO_2} = \frac{\Delta A_{\lambda 2}}{\Delta A_{\lambda 1}} = \frac{ln(I_{n_{\lambda 2}})}{ln(I_{n_{\lambda 1}})} \tag{36}$$

As light passes through the additional optical path $\Delta d$ at systole, from Equation (11), it is written as

$$d_1 = d_2 + \Delta d \tag{37}$$

The normalized intensity of the received light at wavelength $\lambda$ can be expressed as

$$I_{n_\lambda} = \frac{I}{I_{H_{d_2}}} \tag{38}$$

where $I$ represents the light intensity received by the photodetector (PD), and $I_{H_{d2}}$ represents the highest intensity at diastole. The absorbance at wavelength $\lambda$ can be found using the concentrations of oxyhemoglobin and deoxyhemoglobin as follows:

$$\Delta A_\lambda = (\varepsilon_{HbO}(\lambda) \times c_{HbO} + \varepsilon_{HHb}(\lambda) \times c_{HHb}) \times \Delta d \tag{39}$$

Now, replacing $c_{HbO}$ and $c_{HHb}$ in Equation (39) using Equation (4), we get

$$\Delta A_\lambda = (\varepsilon_{HbO}(\lambda) \times SpO_2(c_{HbO} + c_{HHb}) + \varepsilon_{HHb}(\lambda)(1 - SpO_2)(c_{HbO} + c_{HHb})) \times \Delta d$$

$$\text{Or,} \quad \Delta A_\lambda = (\varepsilon_{HbO}(\lambda) \times SpO_2 + \varepsilon_{HHb}(\lambda)(1 - SpO_2)) \times (c_{HbO} + c_{HHb}) \times \Delta d \tag{40}$$

Now, (36) can be expressed as

$$R_{SpO_2} = \frac{\Delta A_{\lambda 2}}{\Delta A_{\lambda 1}} = \frac{(\varepsilon_{HbO}(\lambda 2) \times SpO_2 + \varepsilon_{HHb}(\lambda 2)(1 - SpO_2)) \times (c_{HbO} + c_{HHb})}{(\varepsilon_{HbO}(\lambda 1) \times SpO_2 + \varepsilon_{HHb}(\lambda 1)(1 - SpO_2)) \times (c_{HbO} + c_{HHb})} \tag{41}$$

Finally, the oxygen saturation ($SpO_2$) can be calculated as

$$SpO_2 = \frac{\varepsilon_{HHb}(\lambda 2) - (\varepsilon_{HHb}(\lambda 1) \times R_{SpO_2})}{(\varepsilon_{HHb}(\lambda 2) - \varepsilon_{HbO}(\lambda 2)) + (\varepsilon_{HbO}(\lambda 1) - \varepsilon_{HHb}(\lambda 1) \times R_{SpO_2})} \tag{42}$$

## 3. Methodology

To estimate the HbA1c value noninvasively, in this study, two different ratio calibrations were applied. Each ratio equation used for calibration for $SpO_2$ and HbA1c estimation is defined differently. In the ratio calibration, two XGBoost models were used for calibrating the ratio values for estimating $SpO_2$ and HbA1c. If necessary, value (HbA1c) calibration can be used in addition to ratio calibrations to improve the accuracy. Note that value (HbA1c) calibration can be optionally adopted when the desired performance cannot be achieved only with ratio calibrations.

*3.1 Dataset-Related Information*

To evaluate the accuracy of the model and validity of the theory, we proceeded using the same data from the 20 subjects noted in [14]. Of these, thirteen were pre-diabetic, three were diabetic, and four were normal. The participants ranged in age from 25 to 55 (**31.6 ± 10**) years. Five of the subjects were female, and fifteen were male. The mean and standard deviation (SD) of finger width and BMI in the dataset were 1.30±0.13 and 28.86±3.74, respectively.

In this study, devices such as the Schiller Argus OXM Plus and invasive Bio-Hermes A1C EZ 2.0 were used to collect %$SpO_2$ and data on the National Glycohemoglobin

Standardization Program (NGSP) %HbA1c levels, respectively. The study also involved recording a 4-minute PPG signal, with 2 minutes being transmissive and the remaining 2 minutes being reflective measurements. The transmissive PPG signal was chosen for the study as it aligns with the theoretical derivation used in the research.

The Institutional Review Board (IRB) of Kookmin University in Seoul, Korea, provided guidelines for the study protocol. The IRB procedures of Kookmin University's were followed in conducting this study. Additionally, all participants gave their permission in advance for the data to be used academically. More details on the clinical dataset information can be found in [14]. The statistical summary of the entire dataset used in this study is shown in Figure 2 and Table 3.

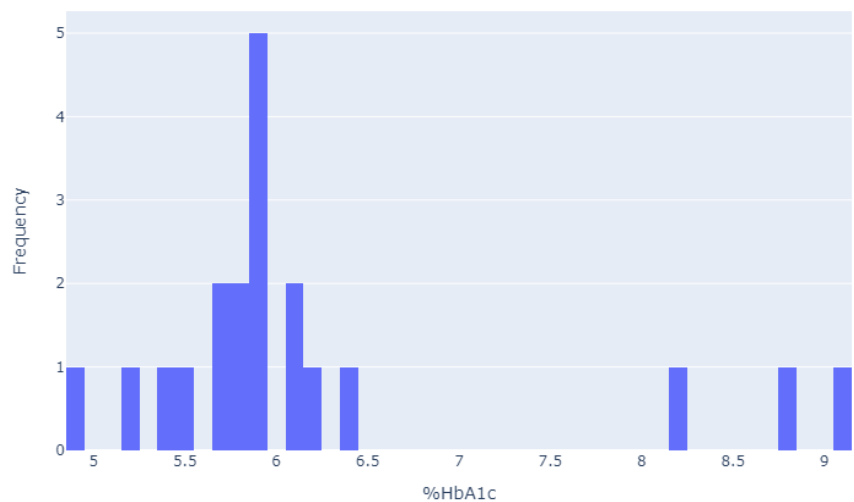

(a)

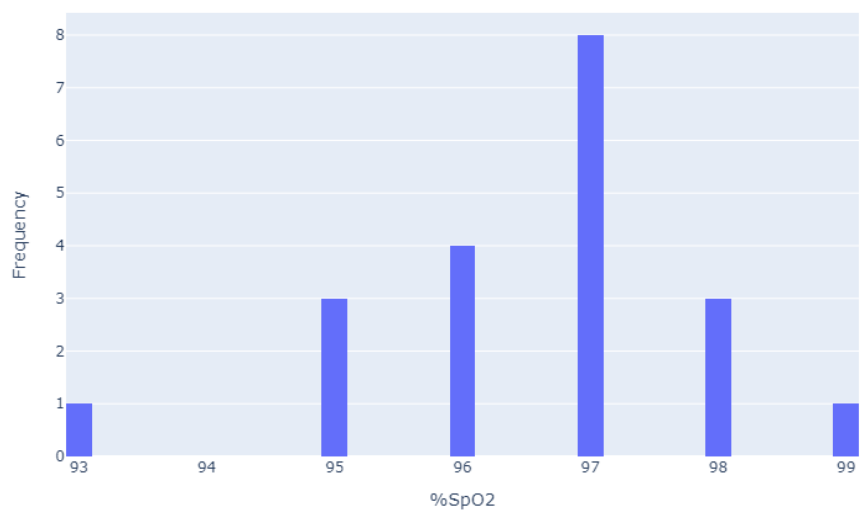

(b)

**Figure 2.** Histograms of the measured dataset: (a) %NGSP HbA1c and (b) %SpO$_2$ values.

**Table 3.** Statistics of the measured %HbA1c and %SpO$_2$ data.

|  | Min | Max | Mean | Median | SD | Variance | 25th Percentile | 75th Percentile |
|---|---|---|---|---|---|---|---|---|
| %HbA1c | 4.9 | 9.1 | 6.224 | 5.9 | 1.0308 | 1.0626 | 5.7 | 6.2 |
| %SpO$_2$ | 93 | 99 | 96.6 | 97.0 | 1.4142 | 2.0 | 96.0 | 97.0 |

*3.2. Proposed Method: HbA1c Estimation Using Only Two Wavelengths*

The workflow diagram of the proposed method is shown in Figure 3.

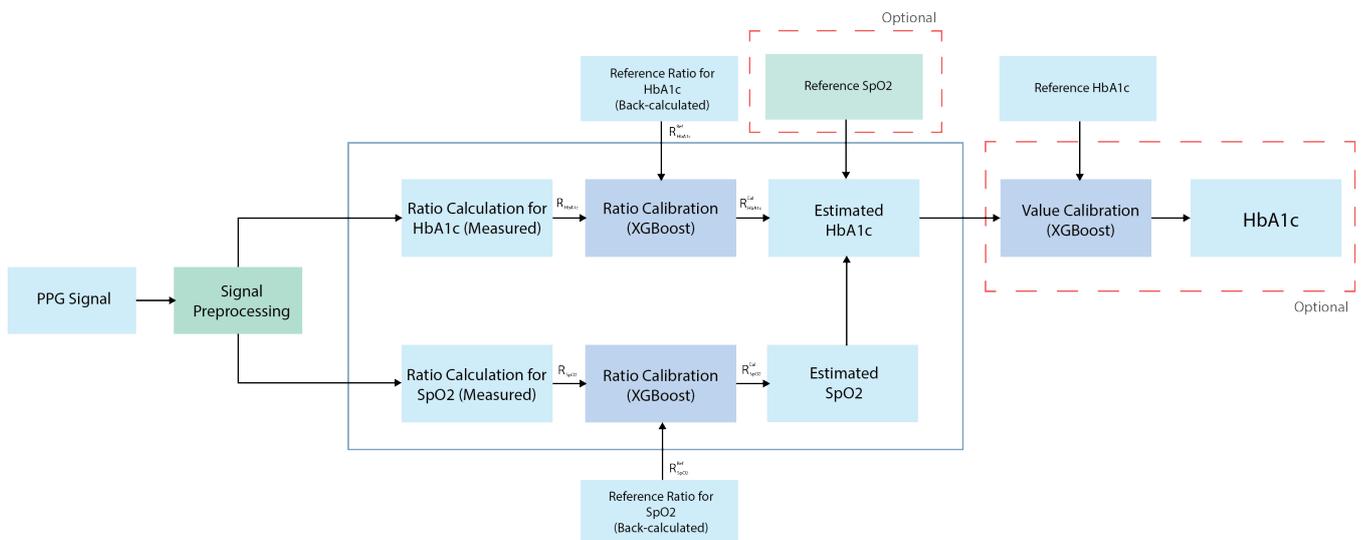

**Figure 3.** HbA1c estimation using only two wavelengths.

Data acquisition and data preprocessing were performed similar to [14]. For data acquisition, we used the commercial sensor module DFRobot SEN0212 comprising a color sensor (TCS34725) and a set of four white LEDs [14]. TCS34725 is a highly sensitive sensor with three wavelengths (465, 525, and 615 nm). Although the device used here can capture both transmissive and reflective signals of all three wavelengths, only the transmissive signals of two wavelengths (525 and 615 nm) are considered in this study. The PPG waveform is then preprocessed by filtering through a second-order Butterworth lowpass filter with a cutoff frequency of 8 Hz. As can be seen in Figure 3, two XGBoost calibration models were used here for ratio calibrations of SpO$_2$ and HbA1c. Note that in [14], the SpO$_2$ and HbA1c estimates were obtained by simultaneously performing two ratio calibrations using three wavelengths, whereas in this study, the different ratio calibrations were separately performed using only two wavelengths to obtain the SpO$_2$ and HbA1c estimates sequentially.

*3.3. Calibration*

Calibration was performed after dataset creation and data preprocessing. For that reason, first the ratio values are calculated directly from the input intensity values. The ratio value for SpO$_2$ was calculated using Equation (36) from the input PPG signals. Ratio values for HbA1c were calculated using Equation (19) and Equation (34) for blood-vessel and whole-finger model, respectively. Then, the reference ratio values for SpO$_2$ were inversely calculated from the reference SpO$_2$ values using Equation (42). The reference ratio values for HbA1c were also inversely calculated in a similar way using Equation (23) and Equation (35) for blood-vessel model and whole-finger model, respectively. This process of calibrating the ratio values is essential because different people have different finger widths and skin and fat layer qualities. This calibration method is used to reduce the effects of skin, fat layer, and finger width on the PPG signal amplitude. When performing the calibration, the inputs are the ratio values calculated directly from the PPG signal and the corresponding finger width and BMI, while the target (reference) values are inversely calculated from the reference HbA1c values. The calibrated ratio values are then used to estimate the $SpO_2$ and HbA1c values. Here, to obtain the HbA1c estimated value, the estimated $SpO_2$ value is applied, and if the reference $SpO_2$ value is available, this reference value can be used instead of the $SpO_2$ estimate. After ratio calibrations, additional value (HbA1c) calibration can be conducted if necessary. In the value calibration, the HbA1c value estimated from the ratio calibration model may be further calibrated to improve the performance. Finger width and BMI are also considered as features in this case.

For the calibration, the XGBoost model was used. The leave-one-out cross validation (LOOCV) approach was implemented to evaluate the calibration results. LOOCV is a technique for evaluating a machine-learning model's performance. In the LOOCV, the model is trained on all but one of the data points before being evaluated on the remaining data point. This procedure is repeated for each data point, with each point serving as the test set exactly once. The model's overall performance is then determined by averaging the performances of all iterations. LOOCV is a special case of the k-fold cross-validation, where k is equal to the number of data points. As we focus on estimating HbA1c from PPG signals using the Beer–Lambert based model, implementing LOOCV in regression is reliable and unbiased for achieving the desired model performance.

## 4. Results and Discussion

### 4.1. Blood-Vessel Model

After performing ratio calibrations, Clarke's error-grid analysis (EGA) [20, 21] and Bland–Altman analysis plots were used for performance analysis of the estimated HbA1c values. As seen in Figure 4, from the EGA, Zone A contains 15 samples (75%; clinically accurate), Zone B contains 5 samples (25%; data outside of 20% of the reference, but would not lead to inappropriate treatment), and Zone C contains 0 samples (0%; data that would lead to inappropriate treatment). The Bland–Altman analysis indicates that the blood-vessel model provided a bias of − 0.15998 ± 0.961, and the limits of agreement ranged from -1.12 to 0.80. For the estimated %HbA1c values, statistical analysis using mean square error (MSE), mean error (ME), mean absolute deviation (MAD), root mean square error (RMSE), and Pearson's r yielded 0.266, -0.1599, 0.423, 0.5156, and 0.8959, respectively.

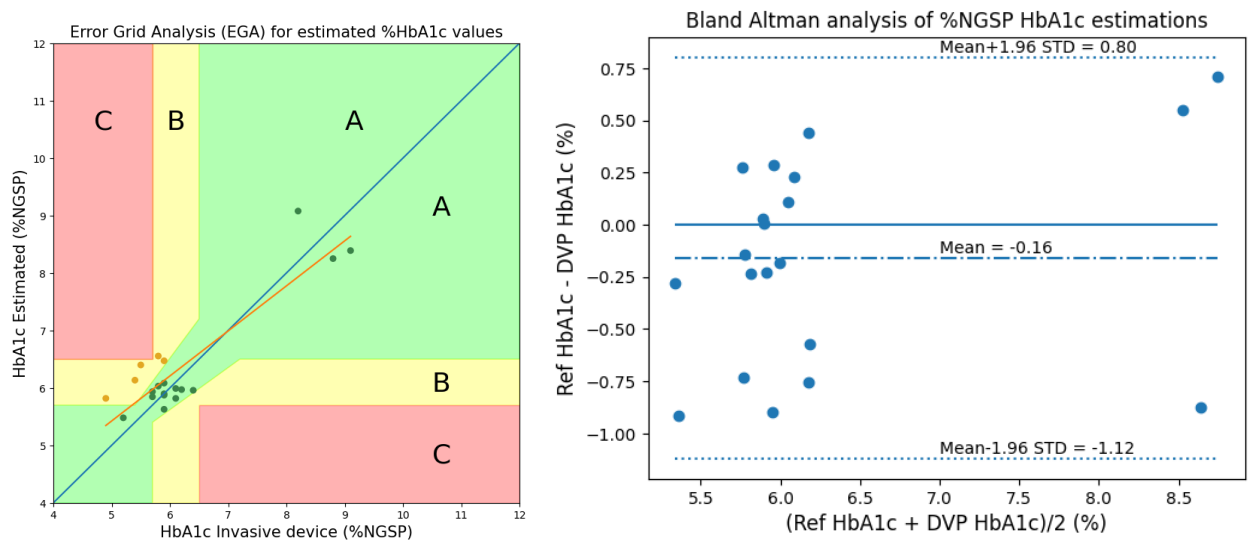

**Figure 4.** EGA and Bland–Altman analysis for HbA1c estimation with blood-vessel model (Considering ratio calibrations only).

If we consider value (HbA1c) calibration in addition to ratio calibrations, the EGA and Bland–Altman analysis results are shown in Figure 5. From the EGA, Zone A contains 17 samples (85%), Zone B contains 3 samples (15%), and Zone C contains 0 samples (0%). The Bland–Altman analysis indicates that the blood-vessel model provided a bias of –0.029 ± 0.8598, and the limits of agreement ranged from -0.89 to 0.83. For the estimated %HbA1c values, statistical analysis using MSE, ME, MAD, RMSE, and Pearson's r yielded 0.259, -0.0118, 0.4366, 0.5087, and 0.8873, respectively.

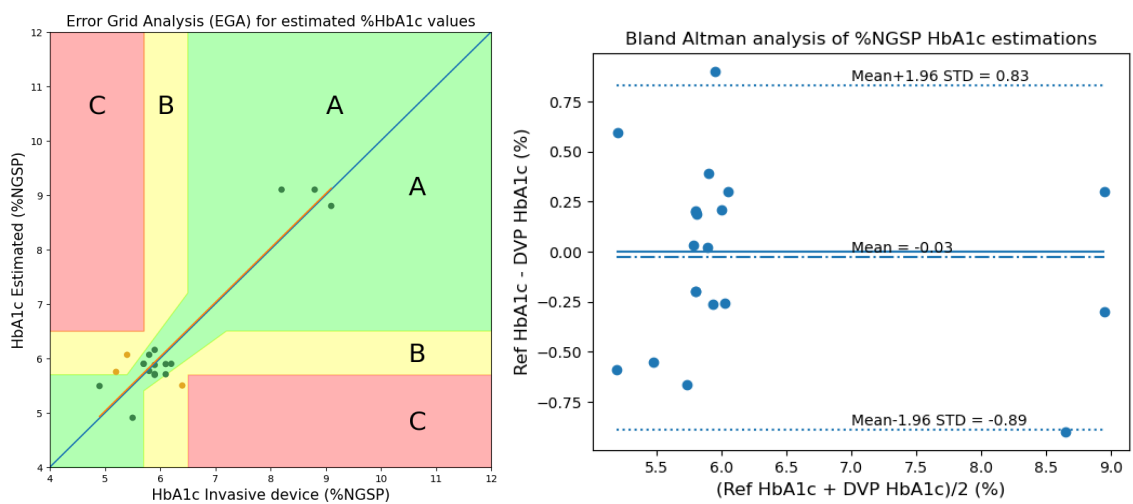

**Figure 5.** EGA and Bland–Altman analysis for HbA1c estimation with blood-vessel model (Considering both ratio and value calibrations).

### 4.2. Whole-Finger Model

After performing ratio calibrations considering the whole-finger model, the results are shown in Figure 6; from the EGA, Zone A contains 16 samples (80%), Zone B contains 4 samples (20%), and Zone C contains 0 samples (0%). The Bland–Altman analysis indicates that the bias was – 0.0718 ± 0.9260, and the limits of agreement ranged from –1.00 to 0.85. The limits of agreement of the whole-finger model were smaller than those of the blood-vessel model. For the estimated %HbA1c values, statistical analysis using MSE, ME, MAD, RMSE, and Pearson's r yielded 0.224, -0.0559, 0.3914, 0.4736, and 0.9052, respectively.

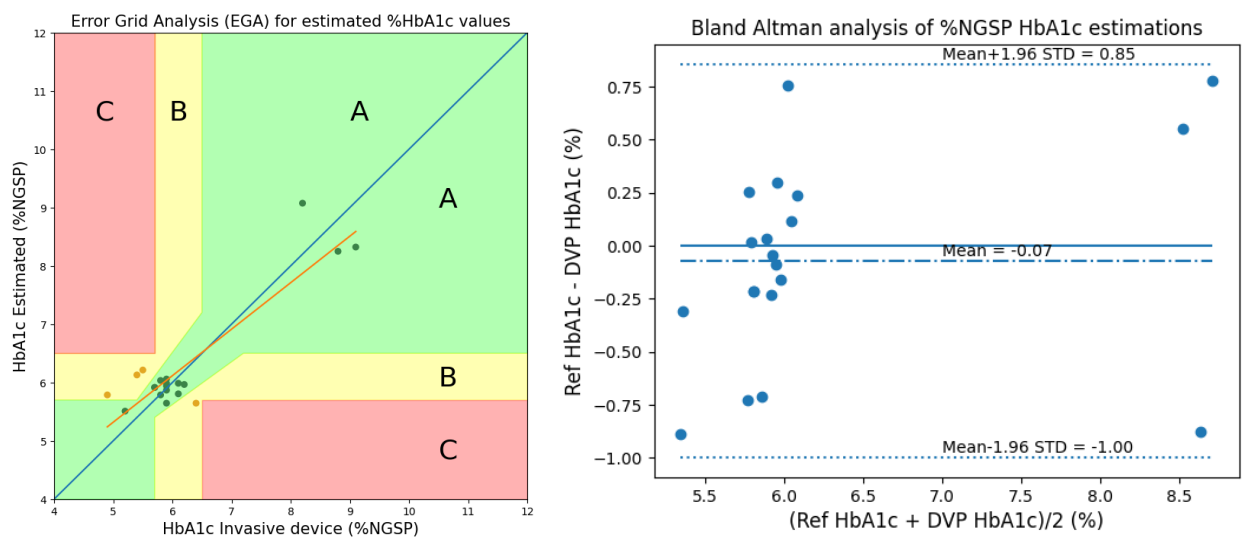

**Figure 6.** EGA and Bland–Altman analysis for HbA1c with whole-finger model (Considering ratio calibrations only).

If we consider value (HbA1c) calibration in addition to ratio calibrations, the EGA and Bland–Altman analysis results are shown in Figure 7. From the EGA, Zone A contains 17 samples (85%), Zone B contains 3 samples (15%), and Zone C contains 0 samples (0%). The Bland–Altman analysis indicates that the bias was 0.0066 ± 0.8623, and the limits of agreement ranged from – 0.86 to 0.87. For the estimated %HbA1c values, statistical analysis using MSE, ME, MAD, RMSE, and Pearson's r yielded 0.194, 0.0066, 0.3662, 0.4400, and 0.9296, respectively.

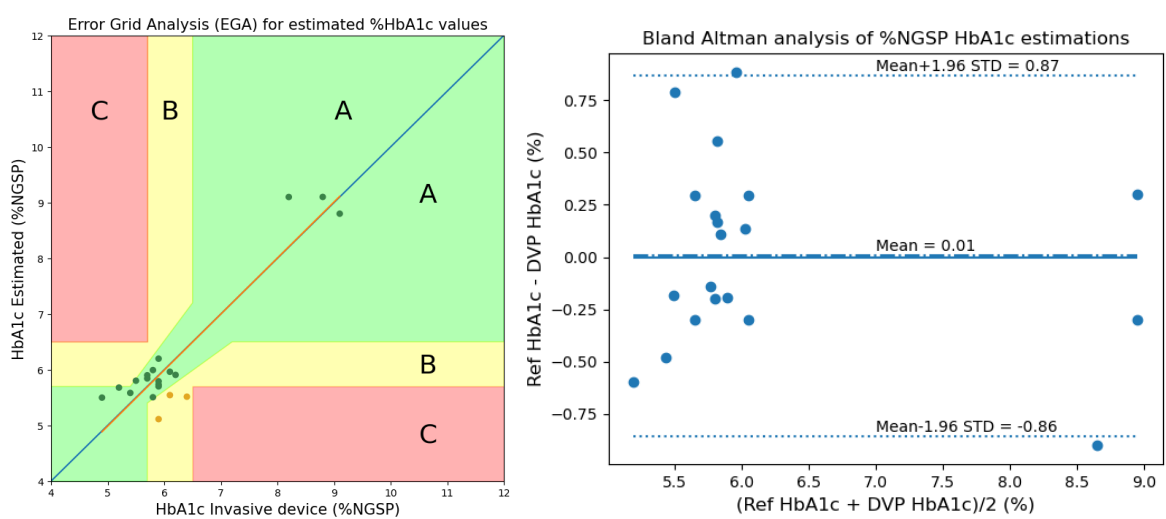

**Figure 7.** EGA and Bland–Altman analysis for HbA1c with whole-finger model (Considering both ratio and value calibrations).

### 4.3. SpO$_2$ Estimation

For the estimated SpO$_2$ obtained through ratio calibration, the scatter plot and Bland–Altman analysis results are plotted in Figure 8. The Bland–Altman analysis provided a bias of – 0.0894 ± 3.293, and the limits of agreement ranged from – 3.38 to 3.20.

For the estimated %SpO$_2$ values, statistical analysis using MSE, ME, MAD, and RMSE are 2.831, -0.089, 1.392, and 1.683, respectively. The reference closeness factor (RCF) defined as Equation (43) was found to be 0.986.

$$RCF = \frac{1}{N}\sum_{i=1}^{N}\left(1 - \frac{\left|SpO_2^{Ref}(i) - SpO_2^{Est}(i)\right|}{100}\right) \qquad (43)$$

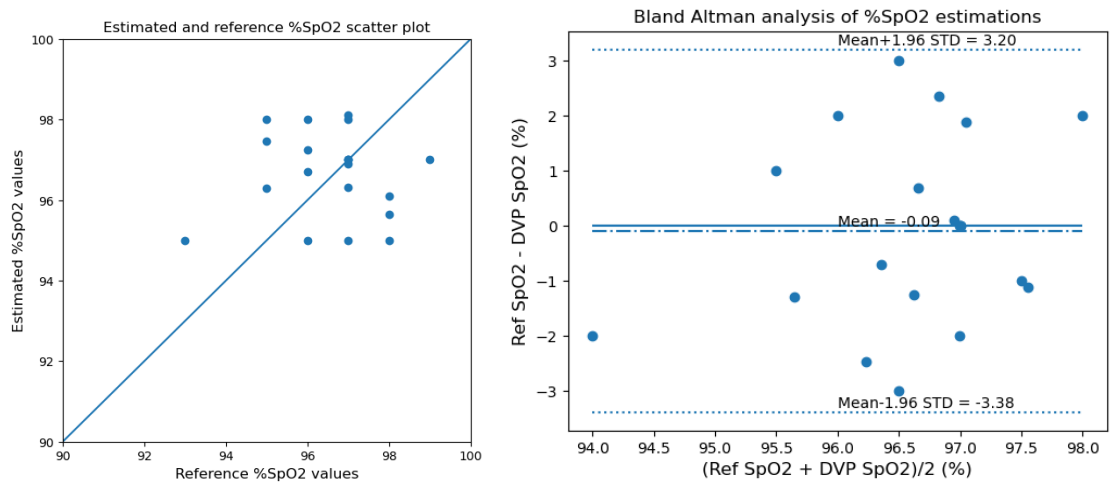

**Figure 8.** Scatter plot and Bland–Altman analysis for the estimated SpO$_2$ values after ratio calibration.

### 4.4. Performance Comparisons Using the Evaluation Metrics

Table 4 shows the performance comparison results for the accuracy of the HbA1c estimates between the previous study [14] using three wavelengths and this study using only two wavelengths. We see that in the HbA1c estimation, even though only two wavelengths were used, the performance was comparable to (slightly worse than) that of the previous study when only ratio calibrations were applied; when value (HbA1c) calibration was applied in addition to ratio calibrations, the performance was almost equal to or slightly better compared to that of the previous study [14].

**Table 4.** HbA1c estimation performance comparison between this study and the previous study.

| Metric<br>Method | MSE | ME | MAD | RMSE | Pearson's r |
| --- | --- | --- | --- | --- | --- |
| Blood-vessel [14] | 0.211 | -0.031 | 0.375 | 0.459 | 0.916 |
| Whole-finger [14] | 0.110 | -0.065 | 0.271 | 0.332 | 0.959 |
| Blood-vessel (Proposed1) | 0.266 | -0.159 | 0.423 | 0.515 | 0.896 |
| Whole-finger (Proposed1) | 0.224 | -0.055 | 0.391 | 0.473 | 0.905 |
| Blood-vessel (Proposed2) | 0.193 | -0.029 | 0.363 | 0.439 | 0.929 |
| Whole-finger (Proposed2) | 0.194 | 0.007 | 0.366 | 0.440 | 0.930 |

\* Proposed1: Using ratio calibrations only, Proposed2: Using both ratio calibrations and value (HbA1c) calibration

Table 5 shows the EGA-based comparison between the previous study [14] for estimating HbA1c using three wavelengths and this study using only two wavelengths. The number inside the table indicates the number of points (%values inside bracket) in the corresponding zone. For the whole-finger model, the performances of Proposed1 and Proposed2 are similar. However, for the blood-vessel model, the performance of Proposed2 is slightly better, with 80% of data points inside zone A. Compared with the previous study [14], Proposed2 method thus performed better.

**Table 5.** EGA-based comparison between this study and the previous study.

| Method \ Zone | A | B | C |
|---|---|---|---|
| Blood-vessel [14] | 15(75) | 5(25) | 0(0) |
| Whole-finger [14] | 18(90) | 2(10) | 0(0) |
| Blood-vessel (Proposed1) | 15(75) | 5(25) | 0(0) |
| Whole-finger (Proposed1) | 17(85) | 3(15) | 0(0) |
| Blood-vessel (Proposed2) | 16(80) | 4(20) | 0(0) |
| Whole-finger (Proposed2) | 17(85) | 3(15) | 0(0) |

* Proposed1: Using ratio calibrations only, Proposed2: Using both ratio calibrations and value (HbA1c) calibration

Table 6 summarizes the Bland–Altman-analysis-based comparison between the previous study [14] for estimating HbA1c using three wavelengths and this study using only two wavelengths. We see that the limits of agreement for the proposed methods are not very different compared to the previous method [14]. The 95% limits of agreement should contain the difference between the estimates and reference for 95% of measurements. In the previous study [14], considering the whole-finger model, only one data point was found outside this limit. Similarly, in the proposed method here, there is only one data point outside the limit.

**Table 6.** Bland-Altman-analysis-based comparison between this study and the previous study for estimating HbA1c.

| Method \ Metric | Bias | Limit of agreement (95%; 1.96 SD) | Data points out of limit of agreement |
|---|---|---|---|
| Blood-vessel [14] | − 0.03 ± 0.458 | -0.93 to 0.87 | 0 |
| Whole-finger [14] | − 0.06 ± 0.326 | -0.70 to 0.57 | 1 |
| Blood-vessel (Proposed1) | − 0.16 ± 0.961 | -1.12 to 0.80 | 0 |
| Whole-finger (Proposed1) | − 0.072 ± 0.926 | -1.00 to 0.85 | 0 |
| Blood-vessel (Proposed2) | − 0.029 ± 0.859 | -0.89 to 0.83 | 1 |
| Whole-finger (Proposed2) | − 0.0066 ± 0.862 | -0.86 to 0.87 | 1 |

* Proposed1: Using ratio calibrations only, Proposed2: Using both ratio calibrations and value (HbA1c) calibration

$SpO_2$ estimation performance comparison using the evaluation metrics is summarized in Table 7. We see from Table 7 that the RCF score of $SpO_2$ estimated in this study is almost the same as that of the previous study (whole-finger). Considering other statistical analyses, we see that the overall performance of $SpO_2$ estimation in this study is slightly better compared to that of the previous study [14].

**Table 7.** $SpO_2$ estimation performance comparison between this study and the previous study.

| Method \ Metric | MSE | ME | MAD | RMSE | RCF |
|---|---|---|---|---|---|
| Previous (blood-vessel) [14] | 4.038 | 0.178 | 1.676 | 2.010 | 0.983 |
| Previous (whole-finger) [14] | 2.924 | -0.246 | 1.395 | 1.710 | 0.986 |
| Proposed | 2.831 | -0.089 | 1.392 | 1.683 | 0.986 |

Table 8 summarizes the comparisons based on Bland–Altman analysis between the previous study [14] and this study in estimating $SpO_2$. From this, it is clear that the proposed method performs well in comparison with the previous study.

**Table 8.** Bland-Altman-analysis-based comparison between this study and the previous study for estimating $SpO_2$.

| Method | Bias | Limit of agreement (95%; 1.96 SD) | Data points out of limit of agreement |
|---|---|---|---|
| Previous (blood-vessel) [14] | − 0.178 ± 2.002 | -3.74 to 4.10 | 0 |
| Previous (whole-finger) [14] | − 0.246 ± 1.690 | -3.56 to 3.07 | 0 |
| Proposed | − 0.0894 ± 3.293 | -3.38 to 3.20 | 0 |

## 5. Conclusion

In this study, we used Beer–Lambert law to estimate HbA1c noninvasively from the fingertip by considering only two wavelengths. In our previous work [14], three wavelengths were used, which makes the system relatively complex. Obtaining nearly the same performance, as seen from Table 4, while reducing complexity is the main contribution of this study. To this end, two ratio calibrations were used to obtain results comparable to those of the previous study. In the ratio calibrations, two XGBoost models were used: one for $SpO_2$ and the other for HbA1c. The Pearson's r values for the estimated HbA1c values were 0.896 and 0.905 considering ratio calibrations with the blood-vessel and whole-finger models, respectively.

When value (HbA1c) calibration was applied in addition to ratio calibrations, we could further improve the performance: the Pearson's r values of the estimated HbA1c levels were 0.929 and 0.930 for the blood-vessel and whole-finger models, respectively. Further, as in the previous study, the whole-finger model performed slightly better than the blood-vessel model, as shown in Table 4. We also show that the RCF score of $SpO_2$ estimated in this study is almost the same as that of the previous study (whole-finger model).

We expect that further studies using larger datasets and deep-learning techniques can improve these results. The performance can also be improved by calibration in a more controlled manner, paying more attention to factors such as light scattering, finger-width variability, and data filtering. We also note that there is considerable potential for further research on the noninvasive estimation of HbA1c.

**Authors' contributions:** M.S.T.: Conceptualization, Methodology, Software, Writing-Original draft; T.-H.K.: Formal analysis; H.-K.K.: Resources, Validation; K.-D.K.: Validation, Supervision, Project administration, Funding acquisition.

**Funding**: This work was supported by the National Research Foundation (NRF) of Korea funded by the Ministry of Science and ICT (2022R1A5A7000765) and was also supported by Basic Science Research Program through the National Research Foundation (NRF) of Korea funded by the Ministry of Education (NRF-2022R1A2C2010298). This research was also supported through the Korea Industrial Technology Association (KOITA) funded by the Ministry of Science and ICT (MSIT).

**Data Availability Statement**: We have created our own dataset for this study. Since further research is underway, we are unable to publish the dataset at present.

**Ethical Statement:** All protocols and procedures in this study were approved by the Institutional Review Board (IRB) of Kookmin University, Seoul, Korea (approval date: 17th July 2020). The procedures followed the Helsinki Declaration of 1975, as revised in 2008. All human participants agreed in advance to participate and share data for academic research purposes. The IRB protocol number is: KMU-202006-HR-237.

**Conflicts of Interest:** The authors declare that they have no conflict of interest. The funders had no role in the design of the study; in the collection, analyses, or interpretation of data; in the writing of the manuscript; or in the publication of the results.

**Appendix A**

The hardware device that we implemented provides three wavelengths (465, 525, and 615 nm), and we would like to provide a rationale for the selection of two out of the three wavelengths. For selecting the proper wavelength pair, $M_h$ and $M_s$ were calculated using Equations (A1) and (A2) according to [21]. The wavelengths used here are 465 nm (blue), 525 nm (green), and 615 nm (red). These results are summarized in Table A1.

$$M_h = \frac{\sum_{i=1}^{2}\frac{dR_i}{dHbA1c}}{2} \tag{A1}$$

$$M_s = \frac{\sum_{i=1}^{2}\frac{dR_i}{dSpO_2}}{2} \tag{A2}$$

**Table A1.** $M_h$ and $M_s$ statistics.

|  | Blood-Vessel Model | | | Whole-Finger Model | | |
| --- | --- | --- | --- | --- | --- | --- |
|  | GR | BR | BG | GR | BR | BG |
| $M_h$ | 0.342204 | 0.45122 | 0.03224 | 0.34758 | 0.39087 | 0.00068 |
| $M_s$ | 1.964076 | 1.95179 | 1.97210 | 1.93640 | 0.19509 | 0.31324 |

* GR: Green-Red; BR: Blue-Red; BG: Blue-Green

The larger $M_h$ and $M_s$ values, the more sensitive each wavelength pair is to changes in its parameters (HbA1c and SpO$_2$). That is why the higher the values of $M_h$ and $M_s$, the better the performance is. Table A1 shows that for the blue-red pair, both models have the highest $M_h$ values, while the $M_s$ values of both models are not good. For the blue-green pair, only the $M_s$ value for blood-vessel model is the best, but the rest are poor. Note that, in the case of the green-red pair, it can be seen that the overall result is relatively good.

Table A2 shows the results of the EGA-based comparison of HbA1c estimation for different wavelength pairs of the Proposed1 method. We can see that the green-red pair performs better than the other pairs since most of the data points were found in zone A for both blood-vessel and whole-finger models.

**Table A2.** EGA-based comparison of HbA1c estimation for different wavelength pairs.

| Wavelength pair | Method \ Zone | A | B | C |
| --- | --- | --- | --- | --- |
| GR | Blood-vessel (Proposed1) | 15(75) | 5(25) | 0(0) |
| | Whole-finger (Proposed1) | 17(85) | 3(15) | 0(0) |
| BR | Blood-vessel (Proposed1) | 14(70) | 6(30) | 0(0) |
| | Whole-finger (Proposed1) | 15(75) | 5(25) | 0(0) |
| BG | Blood-vessel (Proposed1) | 12(60) | 7(35) | 1(5) |
| | Whole-finger (Proposed1) | 14(70) | 6(30) | 0(0) |

**References**

1. Tapp, R.J.; Tikellis, G.; Wong, T.Y.; Harper, C.A.; Zimmet, P.Z.; Shaw, J.E. Longitudinal Association of Glucose Metabolism with Retinopathy Results from the Australian Diabetes Obesity and Lifestyle (AusDiab) Study on Behalf of the Australian Diabetes Obesity and Lifestyle Study Group. 2008, doi:10.2337/dc07-1707.
2. Chen, C.; Xie, Q.; Yang, D.; Xiao, H.; Fu, Y.; Tan, Y.; Yao, S. Recent Advances in Electrochemical Glucose Biosensors: A Review. *RSC Adv.* **2013**, *3*, 4473–4491, doi:10.1039/C2RA22351A.
3. Bandodkar, A.J.; Wang, J. Non-Invasive Wearable Electrochemical Sensors: A Review. *Trends Biotechnol.* **2014**, *32*, 363–371, doi:10.1016/J.TIBTECH.2014.04.005.
4. Sharma, S.; Huang, Z.; Rogers, M.; Boutelle, M.; Cass, A.E.G. Evaluation of a Minimally Invasive Glucose Biosensor for Continuous Tissue Monitoring. *Anal. Bioanal. Chem.* **2016**, *408*, 8427–8435, doi:10.1007/S00216-016-9961-6/FIGURES/8.
5. Kagie, A.; Bishop, D.K.; Burdick, J.; la Belle, J.T.; Dymond, R.; Felder, R.; Wanga, J. Flexible Rolled Thick-Film Miniaturized Flow-Cell for Minimally Invasive Amperometric Sensing. *Electroanalysis* **2008**, *20*, 1610–1614, doi:10.1002/ELAN.200804253.
6. Li, M.; Bo, X.; Mu, Z.; Zhang, Y.; Guo, L. Electrodeposition of Nickel Oxide and Platinum Nanoparticles on Electrochemically Reduced Graphene Oxide Film as a Nonenzymatic Glucose Sensor. *Sens. Actuators B Chem.* **2014**, *192*, 261–268, doi:10.1016/J.SNB.2013.10.140.
7. Mandal, S.; Marie, M.; Kuchuk, A.; Manasreh, M.O.; Benamara, M. Sensitivity Enhancement in an In-Vitro Glucose Sensor Using Gold Nanoelectrode Ensembles. *J. Mater. Sci.: Mater. Electron.* **2017**, *28*, 5452–5459, doi:10.1007/S10854-016-6207-5/METRICS.
8. Little, R.R.; Roberts, W.L. A Review of Variant Hemoglobins Interfering with Hemoglobin A1c Measurement. *J. Diab. Sci. Technol.* **2009**, *3*, 446–451, doi:10.1177/193229680900300307.
9. Jeppsson, J.O.; Kobold, U.; Barr, J.; Finke, A.; Hoelzel, W.; Hoshino, T.; Miedema, K.; Mosca, A.; Mauri, P.; Paroni, R.; et al. Approved IFCC Reference Method for the Measurement of HbA1c in Human Blood. *Clin. Chem. Lab. Med.* **2002**, *40*, 78–89, doi:10.1515/CCLM.2002.016/MACHINEREADABLECITATION/RIS.
10. Mandal, S.; Manasreh, M.O. An In-Vitro Optical Sensor Designed to Estimate Glycated Hemoglobin Levels. *Sensors* **2018**, *18*, 1084, doi:10.3390/S18041084.
11. Martín-Mateos, P.; Dornuf, F.; Duarte, B.; Hils, B.; Moreno-Oyervides, A.; Bonilla-Manrique, O.E.; Larcher, F.; Krozer, V.; Acedo, P. In-Vivo, Non-Invasive Detection of Hyperglycemic States in Animal Models Using Mm-Wave Spectroscopy. *Sci. Rep.* **2016**, *6*, 1–8, doi:10.1038/srep34035.
12. Usman, S.; Bani, N.A.; Kaidi, H.M.; Aris, S.A.M.; Jalil, S.Z.A.; Muhtazaruddin, M.N. Second Derivative and Contour Analysis of PPG for Diabetic Patients. In: *Proceedings of 2018 IEEE EMBS Conference on Biomedical Engineering and Sciences*, **2019**, 59–62, doi:10.1109/IECBES.2018.8626681.
13. Saraoğlu, H.M.; Selvi, A.O. Determination of Glucose and Hba1c Values in Blood from Human Breath by Using Radial Basis Function Neural Network via Electronic Nose. In: *Proceedings of the National Biomedical Engineering Meeting*, **2014**.
14. Hossain, S.; Sengupta, S.; Kwon, T.H.; Kim, K.D. Derivation and Validation of Gray-Box Models to Estimate Noninvasive in-Vivo Percentage Glycated Hemoglobin Using Digital Volume Pulse Waveform. *Sci. Rep.* **2021**, *11*, 1–18, doi:10.1038/s41598-021-91527-2.
15. Yadav, J.; Rani, A.; Singh, V.; Murari, B.M. Prospects and Limitations of Non-Invasive Blood Glucose Monitoring Using near-Infrared Spectroscopy. *Biomed. Signal Process. Control* **2015**, *18*, 214–227, doi:10.1016/J.BSPC.2015.01.005.
16. Wu, Y.; Hossain, S.; Satter, S.; Kwon, T.-H.; Kim, K.-D. Optical Measurement of Molar Absorption Coefficient of HbA1c: Comparison of Theoretical and Experimental Results. *Sensors* **2022**, *22*, 8179, doi:10.3390/S22218179.
17. Nishidate, I.; Yoshida, K.; Kawauchi, S.; Sato, S.; Sato, M. Tabulated Molar Extinction Coefficient for Hemoglobin in Water. *J. Jpn. Soc. Laser Surg. Med.* **1999**, *32*, 394–401, doi:10.2530/JSLSM.32.394.
18. Saidi, I.S. Transcutaneous Optical Measurement of Hyperbilirubinemia in Neonates - ProQuest, Rice University ProQuest Dissertations, **1992**.
19. Segelstein, D.J. The Complex Refractive Index of Water. **1981**.
20. Lochner, C.M.; Khan, Y.; Pierre, A.; Arias, A.C. All-Organic Optoelectronic Sensor for Pulse Oximetry. *Nat. Commun.* **2014**, *5*, 1–7, doi:10.1038/ncomms6745.
21. Hossain, S.; Kim, K.-D. Comparison of Different Wavelengths for Estimating HbA1c and SpO$_2$ Noninvasively Using Beer-Lambert Law and Photon Diffusion Theory Derived Models. *J. Kor. Soc. Commun.* **2021**, *46*, 1301–1308, doi:10.7840/KICS.2021.46.8.1301.